\documentclass[prb,twocolumn,showpacs]{revtex4}

\usepackage{graphicx}

\begin{document}

\title{Effect of the spatial reflection symmetry on the distribution of the
parametric conductance derivative in ballistic chaotic cavities}

\author{Mois\'es Mart\'{\i}nez-Mares}
\affiliation{Departamento de F\'{\i}sica, Universidad Aut\'onoma
Metropolitana-Iztapalapa, Av. San Rafael Atlixco 186, Col. Vicentina, 09340
M\'exico D. F., M\'exico}

\author{E. Casta\~no}
\affiliation{Departamento de F\'{\i}sica, Universidad Aut\'onoma
Metropolitana-Iztapalapa, Av. San Rafael Atlixco 186, Col. Vicentina, 09340
M\'exico D. F., M\'exico}

\date{\today}

\begin{abstract}
We study the effect of left-right symmetry on the distribution of the
parametric derivative of the dimensionless conductance $T$ with respect to an
external parameter $X$, $\partial T/\partial X$, of ballistic chaotic cavities
with two leads, each supporting $N$ propagating modes. We show that $T$ and
$\partial T/\partial X$ are linearly uncorrelated for any $N$. For $N=1$ we
calculate the distribution of $\partial T/\partial X$ in the presence and absence of
time-reversal invariance. In both cases, it has a logarithmic singularity at zero
derivative and algebraic tails with an exponent different from the one of the
asymmetric case. We also obtain explicit analytical results for the mean and
variance of the distribution of $\partial T/\partial X$ for arbitrary $N$.
Numerical simulations are performed for $N=5$ and 10 to show that the
distribution $P(\partial T/\partial X)$ tends towards a Gaussian one when $N$
increases.
\end{abstract}

\pacs{05.45.-a, 42.25.Bs, 73.23.-b}

\maketitle

\section{Introduction}
\label{sec:intro}

In recent years there has been great interest in coherent quantum transport
in mesoscopic devices in both experiment and theory (for a review see Ref.
\cite{Beenakker1997}). In quantum systems whose classical dynamics is
chaotic, phase coherence and quantum interference give rise to sample-to-sample
fluctuations in most transport properties with respect to small perturbations in
incident energy, cavity shape, or an external parameter such as an applied
magnetic field. Therefore an statistical analysis by random-matrix theory (RMT)
becomes applicable; the details of the problem being irrelevant emphasis must be
put on universal aspects such as the symmetries of the system. Then,
time-reversal invariance (TRI) and spin-rotational symmetries \cite{MQP-1994}, or
the spatial symmetries present in the system characterize their universal
statistics \cite{Gopar1996,Baranger1996,Martinez2000,Falko1}.

Some experiments concerned with the parametric dependence of the conductance
study ballistic quantum dots connected by leads with few propagating modes to
electron reservoirs, as happens in semiconductor systems
\cite{Marcus1992,Chang1994,Chan1995,Keller1996,Huibers1998}. The conductance is
measured as a function of the magnetic field and the shape of the quantum dot
(controlled by gate voltages). The derivative of the conductance with respect to an
external parameter, which is analogous to the level velocity
\cite{Simons,Fyodorov,Taniguchi,Fyodorov2} of the dot, is a very important quantity
that gives us the response of a mesoscopic sample to a small perturbation. Only the
asymmetrical situation has been considered in the literature using an RMT approach
\cite{Brouwer1997,Huibers1998,Fyodorov3,Fyodorov4,Fyodorov5}.

Other methods to verify RMT predictions are given by wave scattering experiments,
such as microwaves \cite{Stein1995,Schanze2001,Schanze2004} and acoustic resonators
\cite{Schaadt}. In the microwave technique, for example, the conductance is
measured as a function of the frequency of incidence, applied magnetic field, and
cavity shape.

In these wave scattering experiments the external parameters are easier to control.
The symmetric case can be constructed to analyze the strong effects of spatial
symmetries on the conductance, predicted by RMT \cite{Gopar1996,Baranger1996,
Martinez2000}. Experiments have been conducted on left-right symmetric cavities
\cite{Schanze2004}. However, no theoretical results exist for the distribution of
the conductance velocities. The purpose of this paper is to study the effect of the
spatial left-right symmetry in the distribution of the parametric derivative of the
conductance. Thus, we consider a chaotic left-right symmetric cavity connected to
two symmetrically located leads for two different symmetries, with and without TRI,
in the absence of an applied magnetic field \cite{Mello2004}. Other spatial
symmetries \cite{Baranger1996} are not considered.

When TRI is broken by a magnetic field the problem of left-right symmetric
cavities is mapped to the one of asymmetric cavities in the presence of TRI.
That problem was studied by Brouwer {\it et al} \cite{Brouwer1997} using an RMT
method to calculate analytically the joint distribution of the dimensionless conductance $T$ and
its velocities $\partial T/\partial V$, $\partial T/\partial X$, with respect to
the gate voltage $V$ and a external parameter $X$ (typically a magnetic field),
for a quantum dot with two single-mode point contacts. The statistical
distributions $P(\partial T/\partial V)$, $P(\partial T/\partial X)$ show
algebraic tails, and in the absence (presence) of TRI they have a cusp
(divergence) at zero velocity, and the second moment is finite (infinite). RMT
predictions have been verified by experimental results \cite{Huibers1998}.

In the present work we perform analytical calculations for the one channel
case to reduce to quadratures the distributions of $\partial T/\partial E$,
and the derivative $\partial T/\partial X$ with respect to an external parameter
$X$ (shape deformation parameter in microwave cavities), using a scattering
matrix formalism. By direct numerical integration we obtain
$P(\partial T/\partial E)$, $P(\partial T/\partial X)$. Analytically, we obtain
their mean and variance and show also that $T$, $\partial T/\partial E$,
$\partial T/\partial X$ are linearly uncorrelated for any number of channels.
Besides, we obtain $P(\partial T/\partial E)$, $P(\partial T/\partial X)$ by
numerical simulations using an scattering matrix.

The formal elements such as the scattering matrix for the symmetric case, its
energy and parametric derivatives are introduced in the next section, as
well as the joint distribution of them, for an arbitrary number of propagating
modes $N$ in the leads. The $N=1$ case is the subject of Sect. \ref{subsec:N1}
for both in the presence and absence of TRI. Sect. \ref{subsec:Narb} is
dedicated to an arbitrary $N$ case. Finally, the conclusions are presented in
Sect. \ref{sec:conclu}.


\section{Chaotic scattering by reflection symmetric cavities}
\label{sec:theory-symm}

\begin{figure}
\includegraphics[width=5.0cm]{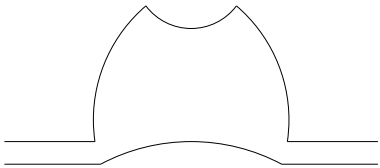}
\caption{A ballistic chaotic left-right symmetric cavity connected to two leads
supporting one channel each one.}
\label{fig:scavity}
\end{figure}

The scattering problem of a ballistic cavity with left-right (LR) symmetry
connected to two leads, each with $N$ transverse propagating modes is described
by the $2N\times 2N$ scattering matrix $S$, which in the stationary case relates
the amplitudes of the outgoing to the incoming plane waves \cite{Newton1982}.
For a system with reflection symmetry (see Fig. \ref{fig:scavity}) the $S$
matrix is block diagonal in a basis of definite parity with respect to
reflections. The general structure for $S$ is \cite{Baranger1996}
\begin{equation} \label{Symmstruct}
S = \left(
\begin{array}{cc}
r & t \\ t & r
\end{array}
\right) ,
\end{equation}
where $r$, $t$ are the $N\times N$ reflection and transmission matrices.

The transmission coefficient, or spinless dimensionless conductance, is
obtained from the $S$ matrix
\begin{equation}\label{Tdef}
T = {\rm tr}(tt^{\dagger});
\end{equation}
it is proportional to the conductance of the cavity,
\begin{equation}
G = (e^2/h)T.
\end{equation}

Matrices with the structure (\ref{Symmstruct}) can be brought to the
block-diagonal form
\begin{equation} \label{block-diagonal}
S = R_0^T \left(
\begin{array}{cc}
S_1 & 0 \\ 0 & S_2
\end{array}
\right) R_0 \, ,
\end{equation}
where $R_0$ is the rotation matrix
\begin{equation} \label{rotation}
R_0 = \frac{1}{\sqrt{2}} \left(
\begin{array}{rr}
\openone_{N} & \openone_{N} \\ -\openone_{N} & \openone_{N}
\end{array}
\right) ,
\end{equation}
$\openone_n$ denotes the $n\times n$ unit matrix and
\begin{eqnarray}
r & = & (S_1 + S_2)/2 \\
t & = & (S_1 - S_2)/2 \label{t-ampl}.
\end{eqnarray}
Here, $S_1=r+t$, $S_2=r-t$ are the most general $N\times N$ scattering matrices.

In the absence of any symmetry the condition of flux conservation implies
unitarity for $S_j$ ($j=1,2$): $S_jS_j^{\dagger}=\openone_N$. In the Dyson's
scheme \cite{Dyson1962} this case is called ''unitary'' and it is designated
by $\beta=2$. In addition, in the presence of time reversal invariance $S_j$
is symmetric: $S_j=S_j^T$. This is the ``orthogonal'' case, designated as
$\beta=1$. It means that two independent $N\times N$ scattering matrices $S_1$,
$S_2$ unitary (and symmetric) for $\beta=2\,(1)$ generate the most general
$2N\times 2N$ unitary (and symmetric) $S$ matrix with the structure
(\ref{Symmstruct}):
\begin{eqnarray}
S S^{\dagger} = \openone_{2N} , \,\, && \mbox{for} \quad \beta=2
\label{unitary} \\
S = S^T , \qquad && \mbox{for} \quad \beta=1 . \label{unitary-symmetric}
\end{eqnarray}

When the classical dynamics of the system is chaotic, most transport
properties are sample specific and a statistical analysis of the
quantum-mechanical problem is called for. That analysis is performed by the
construction of ensembles of physical systems, described mathematically by
ensembles of $S$ matrices distributed according to a probability law. The
starting point is a uniform distribution where the scattering matrix is a
member of one of the {\it circular ensembles}, defined by the {\it invariant
measure} which is the precise formulation of the intuitive notion of {\it
equal a priori probabilities} in the space of scattering matrices.
In our case $S_1$ and $S_2$ are independent of each other, statistically
uncorrelated, and distributed according to the invariant measure defined by
\begin{equation} \label{inv-measure}
d\mu^{(\beta)}(S_j) = d\mu^{(\beta)}(U_0S_jV_0), \qquad (j=1,2).
\end{equation}
Here, $U_0$, $V_0$ are arbitrary but fixed unitary matrices for $\beta=2$,
while $V_0=U_0^T$ for $\beta=1$. Equation (\ref{inv-measure}) defines the
circular unitary (orthogonal) ensemble, CUE (COE), for $\beta=2$ ($\beta=1$)
\cite{Mello1995}.

Then, the invariant measure for $S$ with the structure (\ref{Symmstruct})
can be written as \cite{Gopar1996,Baranger1996}
\begin{equation} \label{inv-meas-symm}
d{\widehat \mu}^{(\beta)}(S) = d\mu^{(\beta)}(S_1) d\mu^{(\beta)}(S_2).
\end{equation}

$S$ matrices of the form given by Eq. (\ref{Symmstruct}), which satisfy Eq. 
(\ref{unitary}) are appropriate for systems with reflection symmetry in the
absence of TRI. With the additional condition (\ref{unitary-symmetric}) it is
appropriate for LR-systems in the presence of TRI. However, when TRI is broken
by a uniform magnetic field, the problem of LR-symmetric cavities is mapped to
one of asymmetric cavities \cite{Baranger1996} with $\beta=1$ with $t$ replaced
by $r$. We are interested in the case when TRI is not broken by a magnetic field
\cite{Mello2004}. Other spatial symmetries \cite{Baranger1996} will not be
considered here.

The parametric derivative of $S_j$ ($j=1,2$) with respect to the energy of
incidence can be defined in terms of a symmetrized form of the Wigner-Smith time
delay matrix, whose eigenvalues are also the proper delay times. The derivative
of $S$ with respect to another external parameter $X$ is defined in similar way.
We have \cite{BrouwerPRL1997}
\begin{equation}\label{dqS}
\frac{\partial S_j}{\partial E} = i \, S_j^{1/2} \, {Q_j}_E \, S_j^{1/2},
\quad
\frac{\partial S_j}{\partial X} = i \, S_j^{1/2} \, {Q_j}_X \, S_j^{1/2} .
\end{equation}
${Q_j}_E$, ${Q_j}_X$ are $N\times N$ Hermitian matrices for $\beta=2$, real
symmetric for $\beta=1$. The eigenvalues of ${Q_j}_E$ are $\hbar^{-1}$ times the
proper delay times.

For classically chaotic cavities the joint distribution of $S_j$, ${Q_j}_E$,
${Q_j}_X$ is \cite{BrouwerPRL1997}
\begin{eqnarray}\label{P(S,QE,QX)}
P_j^{(\beta)}\left(S_j, {Q_j}_E, {Q_j}_X \right) \propto
\left( \det {Q_j}_E \right)^{-2\beta N-3(1-\beta/2)} \nonumber \\
\times
\exp \left\{ -\beta \; {\rm tr}
\left[ \frac{\pi}{\Delta} {Q_j}_E^{-1} +
\left( \frac{\pi X_{\beta}}{2\Delta} {Q_j}_E^{-1}
{Q_j}_X \right)^2 \right] \right\} ,
\end{eqnarray}
where $\Delta$ is the mean level spacing, and $X_{\beta}$ is a typical scale for
$X$ which is not a universal quantity. $S_j$ is independent of ${Q_j}_E$ and
${Q_j}_X$ and it is uniformly distributed in the space of scattering matrices, i.e.
according to the invariant measure $d\mu^{(\beta)}(S_j)$. ${Q_j}_X$ is distributed
as a Gaussian with a width set by ${Q_j}_E$. The rates $\{{x_j}_n=1/{\tau_j}_n\}$
($n=1,\ldots,N$), where the dimensionless time delay ${\tau_j}_n$ is $\Delta/2\pi$
times the $n$th eigenvalue of ${Q_j}_E$, are distributed according to the Laguerre
ensemble, namely \cite{BrouwerPRL1997}
\begin{equation}\label{Laguerre}
P_{\beta} \left( \{ {x_j}_n \} \right) \propto
\prod_{a<b} \left| {x_j}_a - {x_j}_b \right|^{\beta}
\prod_c {x_j}_c^{\beta N/2} e^{-\beta {x_j}_c/2} \, .
\end{equation}
For $N=1$ it reproduces the result of Refs.
\cite{GoparPRL1996,Fyodorov3,Fyodorov5}.

Let denote by $q$ any of the two parameters $E$ or $X$. A convenient
parametrization for $S_j$ and $\partial S_j/\partial q$ ($q=E,\,X$) is
\begin{equation}\label{paraSdqS}
S_j = U_j V_j , \qquad
\frac{\partial S_j}{\partial q} = i\, U_j {Q_j}_q V_j ,
\end{equation}
where $U_j$, $V_j$ are the most general $N\times N$ unitary matrices for
$\beta=2$, while $V_j=U_j^T$ for $\beta=1$.

We parametrize ${Q_j}_X$ and ${Q_j}_E$ as \cite{BrouwerPRL1997}
\begin{eqnarray}
{Q_j}_X & = & \frac 1{X_{\beta}}
{\Psi_j^{-1}}^{\dagger} K_j \Psi_j^{-1}, \label{paraQX} \\
{Q_j}_E & = & \frac{2\pi}{\Delta} {\Psi_j^{-1}}^{\dagger} \Psi_j^{-1}
\label{paraQE}
\end{eqnarray}
where $K_j$ is a $N\times N$ Hermitian matrix for $\beta=2$, real symmetric for
$\beta=1$, that have a Gaussian distribution with zero mean and variance
\begin{equation}\label{varK}
\left\langle (K_j)_{ab} (K_j)_{cd} \right\rangle = \left\{ \begin{array}{ll}
4 \, \delta_{ad} \delta_{bc} & \beta=2 \\
4 \left( \delta_{ad} \delta_{bc} + \delta_{ac} \delta_{bd} \right) &
\beta=1 \end{array} \right. \, ,
\end{equation}
as can be seen by substitution of Eqs. (\ref{paraQX}), (\ref{paraQE}) into
Eq. (\ref{P(S,QE,QX)}). $\Psi$ is a $N\times N$ matrix, complex in the unitary
case, real in the orthogonal one. We define
$Q_j={\Psi_j^{-1}}^{\dagger} \Psi_j^{-1}$ and diagonalize it,
\begin{equation}\label{Qdiag}
Q_j = W_j\hat{\tau_j}W_j^{\dagger}.
\end{equation}
The reciprocals of the eigenvalues $\{{\tau_j}_n\}$ of $\hat{\tau_j}$ are
distributed according to Eq. (\ref{Laguerre}). The matrix of eigenvectors, $W$,
is uniformly distributed in the unitary (orthogonal) group for $\beta=2$
($\beta=1$).


\section{Parametric conductance velocity distributions}


\subsection{The $N=1$ case}
\label{subsec:N1}

In this section we calculate the distributions of the partial derivatives of
the dimensionless conductance, namely $\partial T/\partial E$ and
$\partial T/\partial X$, when the leads support one propagating mode.

For $N=1$ the $S$ matrix is $2\times 2$ and $S_j=e^{i\theta_j}$ ($j=1,2$) is
uniformly distributed in the unit circle, i.e.
$d\mu^{(\beta)}(S_j)=d\theta_j/2\pi$; $S$ is distributed as Eq.
(\ref{inv-meas-symm})
\begin{equation}
d\widehat{\mu}^{(\beta)}(S) = \frac{d\theta_1}{2\pi} \frac{d\theta_2}{2\pi}.
\end{equation}
The transmission coefficient is
\begin{equation} \label{transmission}
T = \frac 12 [1 - \cos(\theta_1 - \theta_2) ] ,
\end{equation}
whose probability distribution is given by \cite{Gopar1996}
\begin{equation} \label{P(T)}
w_{\beta}(T) = \frac 1{\pi\sqrt{T(1-T)}} ,
\end{equation}
for any $\beta$.

The partial derivative with respect to the external parameter $q$ ($q=E,\, X$)
is
\begin{equation} \label{dqT}
\frac{\partial T}{\partial q} = \frac 12 \sin(\theta_1-\theta_2)
\left( \frac{\partial\theta_1}{\partial q} -
\frac{\partial\theta_2}{\partial q} \right) ;
\end{equation}
from (\ref{transmission}) we rewrite the last equation as
\begin{equation}\label{dT(T)}
\frac{1}{\sigma\sqrt{T(1-T)}} \frac{\partial T}{\partial q} =
\left( \frac{\partial\theta_1}{\partial q} -
\frac{\partial\theta_2}{\partial q} \right) ,
\end{equation}
where $\sigma=\pm 1$ depending on the phases of the $1\times 1$ $S_1$,
$S_2$ matrices, $\theta_1$, $\theta_2$. A similar expression to Eq. (\ref{dT(T)})
holds for asymmetric cavities \cite{Brouwer1997} where $\partial T/\partial E$
depends on $T$ but the ratio $(\partial T/\partial E)/\sqrt{T(1-T)}$ is
independent of $T$. However, $T$ is linearly uncorrelated with
$\partial T/\partial q$ ($q=E$, $X$) (see Sect. \ref{subsec:Narb}).

We write Eq. (\ref{dqT}) as
\begin{eqnarray}
\frac{\partial T}{\partial q} = \frac{z_q}2 \sin(\theta_1-\theta_2),
\end{eqnarray}
where we have defined
\begin{equation}
z_q=\frac{\partial\theta_1}{\partial q}-\frac{\partial\theta_2}{\partial q}.
\end{equation}

From Eq. (\ref{dqS}), $\partial\theta_j/\partial E=\tau_j$. In similar way we
denote by $\rho_j=\partial\theta_j/\partial X$. Then,
\begin{equation}
z_E = \tau_1 - \tau_2 \, , \, \quad \, z_X = \rho_1 - \rho_2.
\end{equation}
The sets ($\theta_1$, $\tau_1$, $\rho_1$), ($\theta_2$, $\tau_2$, $\rho_2$)
are statistically independent, while the joint distribution of $\theta_j$,
$\tau_j$, $\rho_j$ ($j=1,2$) is [Eq. (\ref{P(S,QE,QX)})]
\begin{equation} \label{P(theta,tau,rho)}
P_j^{(\beta)} (\theta_j, \tau_j, \rho_j ) = C_{\beta}
\tau_j^{-3-\beta/2}
\exp \left( -\frac{\alpha_{\beta}}{\tau_j} +
\frac{\alpha_{\beta}^2 X_{\beta}^2}{4\beta} \frac{\rho_j^2}{\tau_j^2} \right),
\end{equation}
where $\alpha_{\beta}=\beta\pi/\Delta$,
$C_{\beta}=\alpha_{\beta}^{(2+\beta/2)}X_{\beta}/2\sqrt{\beta\pi}\,
\Gamma(1+\beta/2)$,
and $\Gamma(x)$ is the gamma function.

The distribution of $\partial T/\partial q$ ($q=E,X$) is
\begin{eqnarray}
P_{\beta}(\partial T/\partial q) & = &
\int_0^{2\pi} \frac{d\theta_1}{2\pi}
\int_0^{2\pi} \frac{d\theta_2}{2\pi}
\int_{-\infty}^{\infty} dz_q P'_{\beta}(z_q) \nonumber \\ & \times &
\delta\left[ \frac{\partial T}{\partial q} -
\frac{z_q}2 \sin(\theta_1 - \theta_2) \right] ,
\end{eqnarray}
which can be written as,
\begin{equation} \label{P-partial1}
P_{\beta}(\partial T/\partial q) =
\frac 2{\pi} \int_0^{\pi} P'_{\beta}
\left( \frac{2}{\sin\theta} \frac{\partial T}{\partial q} \right)
\frac{d\theta}{|\sin\theta|} ,
\end{equation}
or, equivalently,
\begin{equation} \label{P-partial2}
P_{\beta}(\partial T/\partial q) =
\frac 8{\pi} \int_{2|\partial T/\partial q|}^{\infty}
\frac{P'_{\beta}(z_q)}{\sqrt{z_q^2 - 4(\partial T/\partial q)^2}} dz_q ,
\end{equation}
where
\begin{eqnarray}
P'_{\beta}(z_q) & = & \int_0^{2\pi} \frac{d\theta_1}{2\pi}
\int_0^{\infty} d\tau_1 \int_{-\infty}^{\infty} d\rho_1
P_1^{(\beta)}(\theta_1, \tau_1, \rho_1) \nonumber \\ & \times &
\int_0^{2\pi} \frac{d\theta_2}{2\pi}
\int_0^{\infty} d\tau_2 \int_{-\infty}^{\infty} d\rho_2
P_2^{(\beta)}(\theta_2, \tau_2, \rho_2) \nonumber \\ & \times &
\delta[z_q - (\xi_1 - \xi_2)],
\end{eqnarray}
with $\xi=\tau$ or $\rho$. Using Eq. (\ref{P(theta,tau,rho)}) we see that the
integrations with respect to $\theta_1$, $\theta_2$ are equal to 1.
For the particular case of $P_{\beta}(\partial T/\partial E)$ the $\rho$
variables are easily integrated, as well as one of the $\tau$ variables, say
$\tau_2$, due to the delta function factor, thus only remaining to be done the
integral with respect to $\tau_1$. After this is done and using Eq.
(\ref{P-partial1}) we get that
\begin{eqnarray} \label{P(dT/dE)n}
P_{\beta} ( \partial T/\partial E ) = A_{\beta}
\left| \frac{\partial T}{\partial E} \right|^{-(2+\beta)}
\int_0^{\pi} d\theta \, (\sin\theta)^{(1+\beta/2)} \nonumber \\ \times
\int_0^{\infty} dv
\frac{ e^{-v}
\left[ \left| \frac{\partial T}{\partial E} \right| \frac{v}{\alpha_{\beta}} +
f_{\beta}(v,\theta,\frac{\partial T}{\partial E}) -
\sin\theta \right]^{2+\beta} }
{ f_{\beta}(v,\theta,\frac{\partial T}{\partial E})
\left[ \left| \frac{\partial T}{\partial E} \right| \frac{v}{\alpha_{\beta}} +
f_{\beta}(v,\theta,\frac{\partial T}{\partial E})\right]^{1+\beta/2} } ,
\nonumber \\
\end{eqnarray}
where $f_{\beta}(v,\theta,\partial T/\partial E)=\{[(\partial T/\partial E)v/
\alpha_{\beta}]^2+\sin^2\theta\}^{1/2}$ and
$A_{\beta}=\alpha_{\beta}^{(1+\beta)}/8\beta\pi[\Gamma(1+\beta/2)]^2$.

For $P(\partial T/\partial X)$ one of the $\rho$'s, say $\rho_2$, is easily
integrated due to the delta function factor, giving a Gaussian integral for
$\rho_1$ which is also easily obtained. The result is substituted in Eq.
(\ref{P-partial2}), the order of integration is interchanged to first perform
the $z_X$ integration, using an integral representation of the zero order
modified Bessel function $K_0(x)$. The remaining integrals, with respect to
$\tau_1$, $\tau_2$, are transformed to new variables to obtain that
\begin{eqnarray} \label{P(dT/dX)n}
&&P_{\beta}(\partial T/\partial X) = B_{\beta} \int_0^1 dx
\frac{(1-x)^{(1+\beta/2)}}{x^{(2+\beta/2)}} e^{-\alpha_{\beta}(1-x)/x}
\nonumber \\ & \times &
\int_0^1 dy \frac{(1-y)^{(1+\beta/2)}}
{y^{(2+\beta/2)} \sqrt{(1-x)^2y^2 + x^2(1-y)^2} }
e^{-\alpha_{\beta}(1-y)/y} \nonumber \\ & \times &
\exp \left[ -g_{\beta}(x,y) \left( \frac{\partial T}{\partial X} \right)^2
\right]
K_0 \left[ g_{\beta}(x,y) \left( \frac{\partial T}{\partial X} \right)^2
\right]
\end{eqnarray}
where $B_{\beta}=2X_{\beta}\alpha_{\beta}^{(3+\beta)}/\pi\sqrt{\beta\pi}
[\Gamma(1+\beta/2)]^2$ and
\begin{equation}
g_{\beta}(x,y) = \frac{\alpha_{\beta}^2 X_{\beta}^2}{2\beta}
\frac{(1-x)^2(1-y)^2}{(1-x)^2y^2+x^2(1-y)^2} .
\end{equation}

\begin{figure}
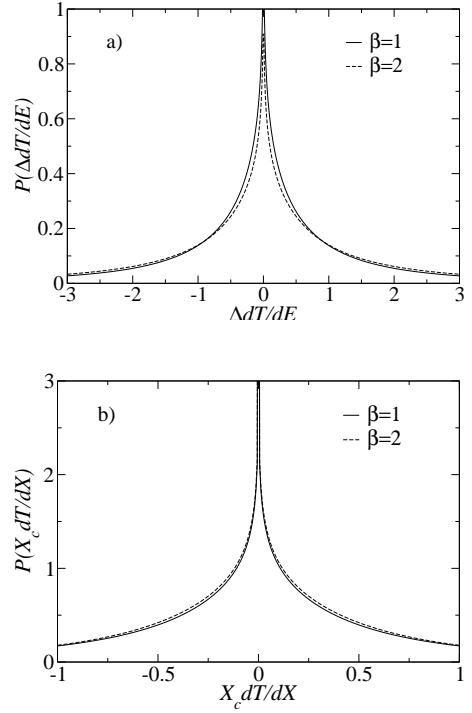

\vskip 0.7cm
\includegraphics[width=6.0cm]{fig2.eps}
\vskip 0.63cm
\includegraphics[width=6.0cm]{fig3.eps}
\caption{Distribution of (a) $\Delta\partial T/\partial E$ and (b)
$X_{\beta}\partial T/\partial X$ for the symmetries $\beta=1,2$, obtained by
numerical integrations of Eqs. (\ref{P(dT/dE)n}) and (\ref{P(dT/dX)n}).}
\label{fig:pdextsn}
\end{figure}

In Fig. \ref{fig:pdextsn} we show the result of the numerical integration of
Eqs. (\ref{P(dT/dE)n}) and (\ref{P(dT/dX)n}). We observe a slight dependence
on $\beta$, while the $T$-distribution is independent of it [see Eq.
(\ref{P(T)})]. In our case, a symmetric one, the dependence on $\beta$ is not as
clearly marked as that of the asymmetric case. Also, in contrast to the
asymmetric case \cite{Brouwer1997}, $P_{\beta}(\partial T/\partial X)$ diverges
logarithmically at zero transmission velocity not only for $\beta=1$ but also for
$\beta=2$.

For large values of $|\partial T/\partial q|$ the tails are algebraic, namely
\begin{equation}
P_{\beta}(\partial T/\partial q) \propto
|\partial T/\partial q |^{-2-\beta/2}
\end{equation}
which again is in contrast with the asymmetric case where the exponent of the
algrabraic tails \cite{Brouwer1997} is $-2-\beta$.


\subsection{Arbitrary $N$}
\label{subsec:Narb}

For arbitrary $N$ we are not able to calculate analytically the whole
distribution of $\partial T/\partial q$ ($q=E$, $X$). Instead, we calculate the
first two moments of the distribution analytically for any $N$ and complement
our results with a numerical simulation of $P_{\beta}(\partial T/\partial q)$
for $N=5$, 10. Also, we calculate the covariance between the energy and
parametric derivatives, as well as the correlation of each one with the
transmission coefficient. The explicit analytical calculations are in the Appendix.

Due to Eq. (\ref{block-diagonal}) the transmission coefficient depends on the
interference term $S_1S_2^{\dagger}$ (see Eq. \ref{Tcoef}), i.e. linearly on
$S_1$ and $S_2$, that are independent from each other, but are equally
distributed random matrices. As a result we have (see the Appendix) 
$\langle T(\partial T/\partial q)\rangle=0$ and
$\langle(\partial T/\partial E)(\partial T/\partial X)\rangle=0$ for all $N$.
This means that $T$, $\partial T/\partial E$, $\partial T/\partial X$ are linearly
uncorrelated variables. However, in general they are correlated (see the  Appendix
\ref{sec:fluc}) as happens in the asymmetric case \cite{Brouwer1997}.

We find (Appendix) that $P_{\beta}(\partial T/\partial q)$ has a zero
mean, i.e. $\langle\partial T/\partial q\rangle=0$, as expected because
$P_{\beta}(\partial T/\partial q)$ is a even function of
$\partial T/\partial q$ (see Fig. \ref{fig:pdextsn}). For $N=1$, which is the
case of Sect. \ref{subsec:N1}, $\langle(\partial T/\partial q)^2\rangle$
diverges in both cases ($\beta=1,2$). For $N>1$ the variance of
$\partial T/\partial X$ is given by
\begin{equation}\label{var(1)}
\left\langle\left( \partial T/\partial X\right)^2\right\rangle =
\frac 1{X_1^2} \frac{ 2( N^2+N+2 )}{(N-2)(N+1)^2}
\end{equation}
for $\beta=1$, and
\begin{equation}\label{var(2)}
\left\langle\left( \partial T/\partial X\right)^2\right\rangle =
\frac 1{X_2^2} \frac{N^2+1}{N(N^2-1)}
\end{equation}
for $\beta=2$; for both cases
\begin{equation}\label{dETSymm}
\left\langle\left(
\partial T/\partial E
\right)^2 \right\rangle =
\frac{\pi^2 X_{\beta}^2}{\Delta^2} \frac{1}N
\left\langle\left(
\partial T/\partial X
\right)^2 \right\rangle
\end{equation}
We see that $\langle(\partial T/\partial q)^2\rangle$ diverges only
for $\beta=1$ when $N=2$.

\begin{figure}
\vskip 0.19cm
\includegraphics[width=8.0cm]{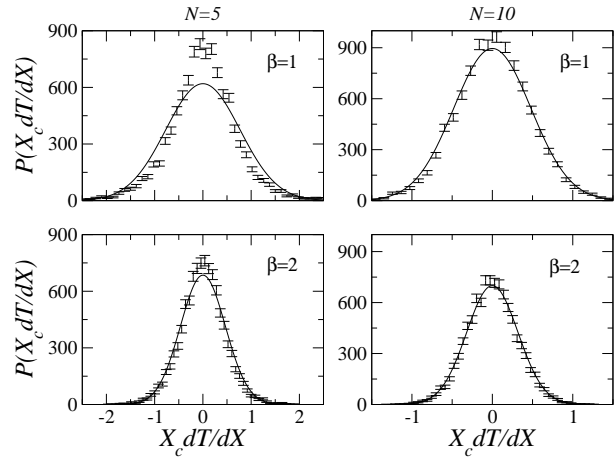}
\caption{Distribution of $\partial T/\partial X$ for $N=5$, 10. When $N$
increases the distribution tends to a Gaussian distribution with variance given
by Eq. (\ref{var(1)}) for $\beta=1$ and Eq. (\ref{var(2)}) for $\beta=2$.}
\label{fig:NarbX}
\end{figure}
\begin{figure}
\vskip 0.19cm
\includegraphics[width=8.0cm]{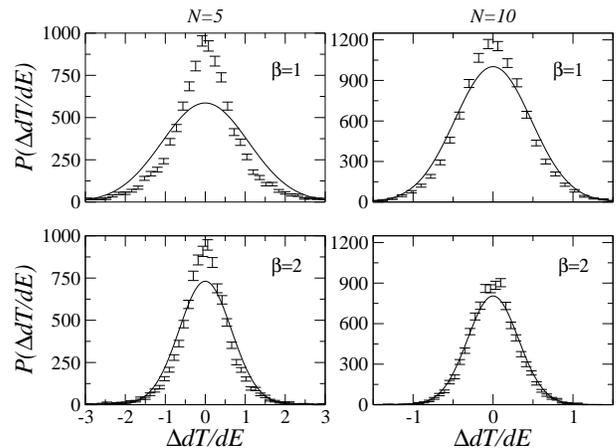}
\caption{As in Fig. {\ref{fig:NarbX}} the distribution of
$\partial T/\partial E$ tends to a Gausssian distribution with variance given by
Eq. (\ref{dETSymm}).}
\label{fig:NarbE}
\end{figure}

For large $N$ cases we resort to numerical simulations. The results for $N=5$,
10 are presented in Fig. \ref{fig:NarbX} for $\partial T/\partial X$, and in
Fig. \ref{fig:NarbE} for $\partial T/\partial E$ where we also include the
Gaussian distributions with variances obtained from our results given by Eqs.
(\ref{var(1)}) and (\ref{var(2)}) for $P_{\beta}(\partial T/\partial X)$, and the
corresponding ones for $P_{\beta}(\partial T/\partial E)$. We observe that the
distribution tends towards a Gaussian as $N$ is increased as expected (see Ref.
\cite{Brouwer1997} and references there in).

As mentioned in a previous publication \cite{Mucciolo}, for chaotic cavities we
implement numerical simulations by generating $10^4$ scattering matrices for each
$S_j$ ($j=1$, 2) given by the Heidelberg approach \cite{Mahaux1969} with $M=100$
poles. Details can be found in Ref. \cite{Mucciolo}.


\section{Conclusions}
\label{sec:conclu}

We have studied the effect of spatial reflection symmetry on the distribution
of the parametric derivative, $\partial T/\partial X$, of the dimensionless
conductance $T$ of a chaotic cavity, with respect to an external parameter $X$.
In particular we considered a left-right ballistic chaotic that has two leads
symmetrically placed, each one supporting $N$ propagating modes, in the
presence (absence) $\beta=1$ ($\beta=2$) of time-reversal invariance (TRI).
The external parameter could be, for instance, the shape of the cavity in the
microwave technique, where our results can be tested \cite{Schanze2004}.

For the particular case $N=1$, we calculated the distribution
$P_{\beta}(\partial T/\partial X)$ by reducing the problem to quadratures and
performing a numerical integration. While the distribution of $T$ does not depend
on $\beta$, the distribution of $\partial T/\partial X$ show a slight dependence
on $\beta$ that is not as clearly marked as in the asymmetric case. Also, in our
case, $P_{\beta}(\partial T/\partial X)$ diverges logarithmically at zero
transmission velocity for any $\beta$. This is in contrast with the asymmetric
case where for $\beta=1$ there is a logarithmic divergence and a cusp $\beta=2$.
We found that $P_{\beta}(\partial T/\partial X)$ has algebraic tails with
an exponent $-2-\beta/2$ different from that in the asymmetric case.

We found that $\partial T/\partial q$ ($q=E,\,X$) depends on $T$ but the ratio
$(\partial T/\partial q)/\sqrt{T(1-T)}$ is $T$-independent as happens in the
asymmetric case. We showed that $T$ and $\partial T/\partial q$ are linearly
uncorrelated for any $N$, although in general they are correlated.

We performed numerical simulations for $N=5,\,10$ to obtain
$P_{\beta}(\partial T/\partial X)$ and compare with a Gaussian distributions with
variances calculated with our analytical result; we see that the distribution
of the parametric derivative of $T$ tends towards a Gaussian distribution when
$N$ becomes large.

Finally, we emphasize that our work is concerned with fully chaotic ballistic
cavities. Our results are consistent with both experimental findings and
simulations \cite{Schanze2004}. It would be also very interesting to single out the
contribution of regular orbits (see \cite{Moura,Backer,Ferry} and references
there in) something that is beyond the scope of the present work, since it would
require to modify the method here employed.


\acknowledgments

MMM thanks C. H. Lewenkopf and A. O. de Almeida for useful comments.
We thank C. H. Lewenkopf for providing a routine on which we based our numerical
simulations. EC thanks his coauthor for introducing him to the the application of
RMT techniques.

\appendix

\section{Mean and variance of $\partial T/\partial q$, ($q=E$, $X$)}
\label{sec:fluc}

From Eqs. (\ref{Tdef}) and (\ref{t-ampl}) we see that
\begin{equation} \label{Tcoef}
T = \frac N2 - \frac 12 \sum_{a,b=1}^N
\text{Re} \, (S_1)_{ab} (S_2^*)_{ab}
\end{equation}
where we have used the unitarity condition for $S_1$ and $S_2$. Because $S_1$ and
$S_2$ are independent but equally distributed random matrices with random phases,
the average of $T$ is $\langle T\rangle=N/2$ for both symmetries
($\beta=1$ and 2), reflecting the spatial symmetry of the cavity.

Let $q$ be one of the two parameter $E$ or $X$ in dimensionless units. For the
same reason as in the last paragraph, the average of the $q$-derivative of T,
\begin{equation} \label{dTcoef}
\frac{\partial T}{\partial q} = -\frac 12 \sum_{a,b=1}^N
\text{Re} \left[ (S_1)_{ab} \frac{\partial (S_2^*)_{ab}}{\partial q} +
\frac{\partial (S_1)_{ab}}{\partial q} (S_2^*)_{ab} \right]
\end{equation}
is $\langle\partial T/\partial q\rangle=0$.

In similar way, after averaging the product of Eqs. (\ref{Tcoef}) and
(\ref{dTcoef}) we get
\begin{eqnarray} \label{TdTcoef}
\left\langle T\frac{\partial T}{\partial q}\right\rangle & = & \frac 14
\sum_{a,b,a',b'=1}^N \text{Re}
\left\langle (S_1)_{ab} (S_1^*)_{a'b'} \right\rangle \nonumber \\  & \times &
\left\langle (S_2^*)_{ab} \frac{\partial (S_2)_{a'b'}}{\partial q}\right\rangle.
\end{eqnarray}
The first term on the right hand side is real positive. For $q=E$, the second
term of Eq. (\ref{TdTcoef}) is pure imaginary as can be seen by substitution of
Eqs. (\ref{paraSdqS}), (\ref{paraQE}) and (\ref{Qdiag}) and averaging over the
unitary matrices $U_2$, $V_2$ and $W_2$ for $\beta=2$, or over $U_2$ and $W_2$
for $\beta=1$. For $q=X$ the parametrization (\ref{paraSdqS}) and Eq.
(\ref{paraQX}) gives a result linear on the matrix $K_2$ which has zero mean.
Therefore, $\langle T(\partial T/\partial q)\rangle=0$. Following the same
procedure we expect that $\langle T^m(\partial T/\partial q)\rangle=0$ with $m$ an
integer (we have verified also the case $m=2$). I.e., $T^m$ is linearly
uncorrelated with $\partial T/\partial E$ and $\partial T/\partial X$. However, in
general they are correlated; for instance we have cheked that
$\langle T(\partial T/\partial q)^2\rangle\neq 0$.

Following the same arguments we get
\begin{eqnarray}\label{vardqT}
& & \left\langle \frac{\partial T}{\partial q}
\frac{\partial T}{\partial q'} \right\rangle =
\text{Re} \, \frac 14 \nonumber \\ & \times & \sum_{a,b,a',b'=1}^N
\left[ \left\langle (S_1)_{ab} (S_1^*)_{a'b'} \right\rangle
\left\langle \frac{\partial (S_2^*)_{ab}}{\partial q}
\frac{\partial (S_2)_{a'b'}}{\partial q'}
\right\rangle \right. \nonumber \\
& + & \left.
\left\langle (S_1)_{ab} \frac{\partial (S_1^*)_{a'b'}}{\partial q'}
\right\rangle \left\langle
\frac{\partial (S_2^*)_{ab}}{\partial q} (S_2)_{a'b'}
\right\rangle
\right] ,
\end{eqnarray}
where $q$ and $q'$ denote $E$ or $X$ independently. Also we have used that $S_1$
and $S_2$ are unitary ($\beta=2$) and symmetric ($\beta=1$) independent and
equally distributed random matrices. When $q=E$ and $q'=X$ the result is
$\langle(\partial T/\partial E)(\partial T/\partial X)\rangle=0$ because the
partial derivative with respect to $X$ is linear in the matrix $K_j$ which has
zero mean. Then, $\partial T/\partial E$ and $\partial T/\partial X$ are
linearly uncorrelated. Again, in general they are correlated. For instance,
$\langle(\partial T/\partial E)^2(\partial T/\partial X)^2\rangle\neq 0$.

Let us consider $q=q'=X$. Using the parametrization (\ref{paraSdqS}) and
averaging over $U_j$ and $V_j$ ($j=1$, 2) for $\beta=2$, or over $U_j$ for
$\beta=1$, using the results of Ref. \cite{Mello1990} we arrive to
\begin{equation}\label{SS*}
\left\langle (S_1)_{ab} (S_1^*)_{a'b'} \right\rangle =
\left\{
\begin{array}{lr}
( \delta_a^{a'} \delta_b^{b'} + \delta_a^{b'}
\delta_b^{a'} )/(N+1) & \beta=1 \\
\delta_a^{a'} \delta_b^{b'}/N & \beta = 2
\end{array}
\right.
\end{equation}
and
\begin{equation} \label{SdXS*}
\left\langle {(S_j)}_{ab}
\frac{\partial {(S_j^*)}_{a'b'}}{\partial X}
\right\rangle = 0 \quad (j = 1,\, 2)
\end{equation}
for both $\beta=1$, and 2, where we used $\langle {Q_X}_{\alpha\alpha}\rangle=0$
because $K_j$ in the parametrization (\ref{paraQX}) has zero mean; while
\begin{equation} \label{dXSdXS*(1)-1}
\left\langle \frac{\partial {(S_2^*)}_{ab}}{\partial X}
\frac{\partial {(S_2)}_{a'b'}}{\partial X} \right\rangle =
\frac{ ( \delta_a^{a'} \delta_b^{b'} + \delta_a^{b'} \delta_b^{a'}) }
{ N ( N+1 ) }
\sum_{\alpha=1}^N \left\langle \left( Q_X^2 \right)_{\alpha\alpha}
\right\rangle
\end{equation}
for $\beta=1$,
\begin{equation} \label{dXSdXS*(2)-1}
\left\langle \frac{\partial {(S_2^*)}_{ab}}{\partial X}
\frac{\partial {(S_2)}_{a'b'}}{\partial X} \right\rangle =
\frac{ \delta_a^{a'} \delta_b^{b'} }{ N^2 }
\sum_{\alpha=1}^N
\left\langle (Q_X^2)_{\alpha\alpha} \right\rangle
\end{equation}
for $\beta=2$. Using Eqs. (\ref{paraQX}), (\ref{paraQE}) and (\ref{varK}),
Eqs. (\ref{dXSdXS*(1)-1}) and (\ref{dXSdXS*(2)-1}) can be written as
\begin{eqnarray} \label{dXSdXS*(1)-2}
\left\langle \frac{\partial {(S_2^*)}_{ab}}{\partial X}
\frac{\partial {(S_2)}_{a'b'}}{\partial X} \right\rangle = 4
\frac{ ( \delta_a^{a'} \delta_b^{b'} + \delta_a^{b'} \delta_b^{a'}) }
{ N ( N+1 ) } \nonumber \\ \times \left[
\sum_{\alpha=1}^N \left\langle \left( Q^2 \right)_{\alpha\alpha} \right\rangle +
\sum_{\alpha,\beta=1}^N \left\langle Q_{\alpha\alpha} Q_{\beta\beta}
\right\rangle \right]
\end{eqnarray}
for $\beta=1$ and
\begin{equation} \label{dXSdXS*(2)-2}
\left\langle \frac{\partial {(S_2^*)}_{ab}}{\partial X}
\frac{\partial {(S_2)}_{a'b'}}{\partial X} \right\rangle = 4
\frac{ \delta_a^{a'} \delta_b^{b'} }{ {N}^2 }
\sum_{\alpha,\beta=1}^N
\left\langle Q_{\alpha\alpha} Q_{\beta\beta}
\right\rangle
\end{equation}
for $\beta=2$.

We substitute Eqs. (\ref{SS*}), (\ref{SdXS*}), (\ref{dXSdXS*(1)-2}) and
(\ref{dXSdXS*(2)-2}) into Eq. (\ref{vardqT}) for $q=q'=X$. After we sum over
$a$, $a'$, $b$, $b'$ the result is
\begin{equation}
\left\langle \left( \partial T/\partial X \right)^2 \right\rangle =
\frac 2{N+1} \left[ \sum_{\alpha=1}^N
\left\langle \left( Q^2 \right)_{\alpha\alpha} \right\rangle +
\sum_{\alpha,\beta}^N
\left\langle Q_{\alpha\alpha} Q_{\beta\beta} \right\rangle
\right]
\end{equation}
for $\beta=1$,
\begin{equation}
\left\langle \left( \partial T/\partial X \right)^2 \right\rangle =
\frac 1N
\sum_{\alpha,\beta}^N
\left\langle Q_{\alpha\alpha} Q_{\beta\beta} \right\rangle
\end{equation}
for $\beta=2$.

Now, the dialgonalization of $Q$, Eq. (\ref{Qdiag}), and the results of Ref.
\cite{Mello1990} lead us to
\begin{equation}\label{vardXT(1)}
\left\langle \left( \partial T/\partial X \right)^2 \right\rangle =
\frac{2N}{N+1} \left[ 2\left\langle\tau_1^2 \right\rangle +
(N-1) \left\langle \tau_1 \tau_2 \right\rangle \right]
\end{equation}
for $\beta=1$ and
\begin{equation}\label{vardXT(2)}
\left\langle \left( \partial T/\partial X \right)^2 \right\rangle =
\left\langle\tau_1^2 \right\rangle +
(N-1) \left\langle \tau_1 \tau_2 \right\rangle
\end{equation}
for $\beta=2$.

Finally, the average with respect to the $\tau$ variables are done using Eq.
(\ref{Laguerre}). When $N=1$ only the first term of Eqs. (\ref{vardXT(1)}) and
(\ref{vardXT(2)}) contribute. But, $\langle\tau_1^2\rangle$ diverges for
$\beta=1$ and $\beta=2$ and then $\langle(\partial T/\partial X)^2\rangle$.
For $N>1$ we have
\begin{equation}
\left\langle\tau_1^2\right\rangle = \left\{
\begin{array}{lr}
2 N!/(N-2)(N+1)! & \beta=1 \\
2 N (N-2)!/(N+1)! & \beta=2
\end{array}
\right.
\end{equation}
and
\begin{equation}
\left\langle\tau_1\tau_2\right\rangle = (N-1)!/(N+1)!
\quad \beta=1,2.
\end{equation}
The final result for $\langle(\partial T/\partial X)^2\rangle$ is given by Eqs.
(\ref{var(1)}) and (\ref{var(2)}) for $\beta=1$ and $\beta=2$ respectively.

In similar way we arrive to the result given by Eq. (\ref{dETSymm}) for the
variance of $\partial T/\partial E$.



\end{document}